\documentclass[iop]{emulateapj}
\usepackage{natbib}
\usepackage{graphicx}
\usepackage{hyperref}
\usepackage{breakurl}
\usepackage{apjfonts}
\usepackage{longtable}
\usepackage{gensymb}
\usepackage{amsmath}
\usepackage{listings}
\newcommand{\msun} {M$_{\sun}$}
\newcommand{\rsun} {R$_{\sun}$}

\newcommand{\Te} {$T_{\rm eff}$}
\newcommand{\logg} {$\log g$}

\begin{document}

\shortauthors{TREMBLAY ET AL}
\shorttitle{3D ATMOSPHERES FOR ELM WDs}

\title{3D Model Atmospheres for Extremely Low-Mass White Dwarfs}

\author{P.-E. Tremblay$^{1,2}$, A. Gianninas$^{3}$, M. Kilic$^{3}$, H.-G. Ludwig$^{4}$, M. Steffen$^{5}$, B. Freytag$^{6}$, and J. J. Hermes$^{7}$}

\affil{$^{1}$Space Telescope Science Institute, 3700 San Martin Drive, Baltimore, MD, 21218, USA; tremblay@stsci.edu} 
\affil{$^{2}$Hubble Fellow}
\affil{$^{3}$Department of Physics and Astronomy, University of Oklahoma, 440 W. Brooks St., Norman, OK, 73019, USA}
\affil{$^{4}$Zentrum f\"ur Astronomie der Universit\"at Heidelberg, Landessternwarte, K\"onigstuhl 12, D-69117 Heidelberg, Germany}
\affil{$^{5}$Leibniz-Institut f\"ur Astrophysik Potsdam, An der Sternwarte 16, D-14482 Potsdam, Germany}
\affil{$^{6}$Department of Physics and Astronomy at Uppsala University, Regementsv\"agen 1, Box 516, SE-75120 Uppsala, Sweden}
\affil{$^{7}$Department of Physics, University of Warwick, Coventry CV4 7AL, UK}

\begin{abstract}
We present an extended grid of mean three-dimensional (3D) spectra for
low-mass, pure-hydrogen atmosphere DA white dwarfs (WDs). We use 
CO5BOLD radiation-hydrodynamics 3D simulations covering
\Te\ =~6000--11,500~K and \logg\ =~5--6.5 ($g$ in cm s$^{-2}$) to derive analytical
functions to convert spectroscopically determined 1D temperatures and
surface gravities to 3D atmospheric parameters. Along with the
previously published 3D models, the 1D to 3D corrections are now available 
for essentially all known convective DA WDs (i.e., \logg\ =~5--9). For low-mass WDs, the correction in
temperature is relatively small (a few per cent at the most), but the
surface gravities measured from the 3D models are lower by as much as
0.35~dex. We revisit the spectroscopic
analysis of the extremely low-mass (ELM) WDs, and demonstrate that the
3D models largely resolve the discrepancies seen in the radius and mass
measurements for relatively cool ELM WDs in eclipsing double WD and WD
+ milli-second pulsar binary systems. We also use the 3D corrections
to revise the boundaries of the ZZ Ceti instability strip, including
the recently found ELM pulsators.
\end{abstract}

\keywords{convection -- hydrodynamics -- white dwarfs}

\section{INTRODUCTION}

Recent three dimensional (3D) radiation hydrodynamical simulations of
the Sun and main-sequence stars have led to
significant improvements in our knowledge of stellar atmospheres and 
spectroscopic abundance measurements
\citep[e.g.][]{asplund09,caffau11,scott15a,scott15b}. The major
differences between the 1D and 3D model atmospheres result from the
insufficient description of convection by the mixing length theory
\citep[][MLT]{bohm58} that is used in the 1D models.

Convection becomes important in hydrogen-atmosphere (DA) white dwarfs (WDs) below about 15,000~K. 
Spectroscopic analyzes of large samples of WDs with 1D model
atmospheres show the so-called ``high \logg\ problem''
\citep{bergeron90,eisenstein06,kepler07,gianninas11, kleinman13}, where the
spectroscopically determined masses of WDs cooler than 13,000~K are
systematically higher than their hotter counterparts by as much as
20\%. There is no known evolution effect or observational bias that 
should lead to such a change in mass for the cooler WDs, and a larger 
average mass is not observed for cool WDs with parallax measurements. In a series of
papers, \citet{tremblay11,tremblay13a,tremblay13b,tremblay13c}
presented the first 3D model atmosphere calculations for DA WDs using
the CO$^{5}$BOLD radiation-hydrodynamics code \citep{freytag12}. The
predicted Balmer line profiles are significantly different in the 3D
models, and the ``high \logg\ problem'' essentially disappears in the
spectroscopic analyzes based on the 3D models.  The previously
published model grid covers the effective temperature range of
6000-15,000 K and the surface gravity range of \logg~=~7--9. To study
the effects of 3D models on the physical parameters of the extremely
low-mass (ELM) WDs with \logg~$<7$, here we extend our 3D model
atmosphere grid down to \logg~=~5.

ELM WDs are helium-core remnants and almost always found in short period binary systems. 
The ELM Survey \citep[][and references therein]{brown10,kilic10a,gianninas14} has discovered
binaries with periods as short as 12--20~min
\citep{brown11,kilic14}. Improving the constraints on the ELM WD
physical parameters, including the spectroscopic mass and
gravitational wave strain, is important for future gravitational wave
missions in the milli-Hertz frequency range. In addition, all of the
known eclipsing and/or tidally distorted double WD systems involve
low-mass WDs. Studying the mass-radius relation for these systems,
\citet{gianninas14} and \citet{kaplan14} find that the 1D
spectroscopic analysis underestimates the radius (or overestimates the
surface gravity) for ELM WDs cooler than about 10,000 K. NLTT 11748 is
an excellent example, where the eclipses constrain the radius and
surface gravity to $R$~=~0.0423--0.0433 \rsun\ and \logg~=~6.32--6.38
\citep{kaplan14}, while the 1D spectroscopic analysis predicts
\logg~=~6.83 \citep{gianninas14}. Therefore, the ``high
\logg\ problem'' seems to impact all convective DA WDs, including ELM
WDs.

We outline the details of our 3D model calculations in Section 2, and
present the 1D to 3D atmospheric parameter correction functions for the 
extended range of $\log{g}=5-9$ in Section 3. We discuss the astrophysical implications,
including the revised parameters for the ELM Survey sample, radius
constraints from the eclipsing and tidally distorted systems as well
as ELM WD + milli-second pulsar binaries, and the ZZ Ceti instability
strip in Section 4. We conclude in Section 5.

\section{3D MODEL ATMOSPHERES}

\begin{figure}
\centering
\includegraphics[scale=0.425,bb=10 167 592 629]{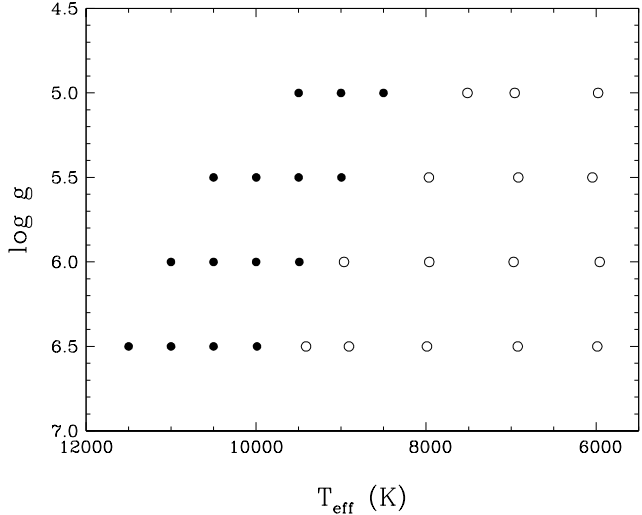}
\figcaption[f_HR.eps]{Surface gravity (logarithmic value) 
  and mean \Te\ for our CO$^{5}$BOLD 3D model atmospheres. The simulations 
	were computed with a bottom boundary layer that is open (open circles) 
	or closed (filled circles) to convective flows. \label{fg:fHR}}
\end{figure}

We have carried out 30 simulations of pure-hydrogen ELM WD
atmospheres with CO$^{5}$BOLD \citep{freytag12}. These 3D simulations
in the range $5.0 \leq \log g \leq 6.5$ provide a direct extension to
the grid presented in \citet{tremblay13c} for $7.0 \leq \log g \leq
9.0$. The calculations also complement the CIFIST grid of CO$^{5}$BOLD
simulations for stars and giants
\citep{cifist,caffau11} with $1.0 \leq \log g \leq
4.5$. Figure~\ref{fg:fHR} illustrates the new ELM
WD 3D simulations in a HR-type diagram. Properties of the
individual models, such as $T_{\rm eff}$ (derived from the temporally
and spatially averaged emergent stellar flux), $\log g$, box
dimensions, and computation time, can be found in Table~\ref{tab:grid}.

\begin{table}[!t]
\caption{Grid of CO$^{5}$BOLD 3D Model Atmospheres for ELM WDs}
\begin{center}
\begin{tabular}{@{}rccccccr@{}l}
\hline
\hline
\noalign{\smallskip}
\multicolumn{1}{c}{\Te} & \logg         & $\log z$ & $\log x$ & $\log \tau_{\rm R,min}$ & $\log \tau_{\rm R,max}$ & $\log t$ & \multicolumn{2}{c}{$\log t_{\rm adv}$} \\
\multicolumn{1}{c}{(K)} & (cm s$^{-2}$) & (cm)     & (cm)     &                         &                         &                                  (s)      & \multicolumn{2}{c}{(s)}               \\

\noalign{\smallskip}
\hline
\noalign{\smallskip}
     5977 &   5.00 &   7.56 &   8.38 &   $-$6.58 &   2.96 &   4.00 &    1&.17  \\
     6958 &   5.00 &   7.71 &   8.46 &   $-$6.08 &   2.99 &   4.00 &    1&.03  \\
     7513 &   5.00 &   7.62 &   8.33 &   $-$5.64 &   2.99 &   4.00 &    0&.92  \\
     8501 &   5.00 &   8.36 &   8.64 &   $-$7.77 &   3.06 &   3.30 &    0&.94  \\
     9001 &   5.00 &   8.38 &   8.64 &   $-$5.46 &   3.00 &   3.30 &    1&.19  \\
     9499 &   5.00 &   8.49 &   8.68 &   $-$8.29 &   3.01 &   3.30 &    1&.54  \\
\noalign{\smallskip}
\hline
\noalign{\smallskip}
     6044 &   5.50 &   7.04 &   7.66 &   $-$6.98 &   2.97 &   3.50 &    0&.75  \\
     6915 &   5.50 &   7.20 &   7.73 &   $-$6.58 &   3.01 &   3.50 &    0&.62  \\
     7967 &   5.50 &   7.37 &   7.98 &   $-$6.57 &   3.02 &   3.50 &    0&.47  \\
     8996 &   5.50 &   7.93 &   8.18 &   $-$7.47 &   3.02 &   2.90 &    0&.46  \\
     9499 &   5.50 &   7.93 &   8.18 &   $-$6.12 &   3.00 &   2.90 &    0&.67  \\
     9998 &   5.50 &   8.00 &   8.22 &   $-$6.75 &   3.01 &   2.90 &    0&.91  \\
    10500 &   5.50 &   8.06 &   8.29 &   $-$7.23 &   3.01 &   2.90 &    1&.40  \\
\noalign{\smallskip}
\hline
\noalign{\smallskip}
     5958 &   6.00 &   6.66 &   7.20 &   $-$7.21 &   2.99 &   3.00 &    0&.38  \\
     6971 &   6.00 &   6.78 &   7.28 &   $-$6.99 &   3.02 &   3.00 &    0&.21  \\
     7962 &   6.00 &   6.89 &   7.34 &   $-$6.82 &   2.99 &   3.00 &    0&.09  \\
     8963 &   6.00 &   7.00 &   7.42 &   $-$5.76 &   2.96 &   3.00 & $-$0&.08  \\
     9491 &   6.00 &   7.32 &   7.65 &   $-$5.07 &   3.00 &   2.40 & $-$0&.03  \\
     9999 &   6.00 &   7.46 &   7.71 &   $-$7.34 &   3.00 &   2.40 &    0&.21  \\
    10499 &   6.00 &   7.52 &   7.75 &   $-$5.75 &   3.01 &   2.40 &    0&.40  \\
    11001 &   6.00 &   7.49 &   7.75 &   $-$5.00 &   2.96 &   2.40 &    0&.77  \\
\noalign{\smallskip}
\hline
\noalign{\smallskip}
     5986 &   6.50 &   6.10 &   6.57 &   $-$7.01 &   2.99 &   2.50 & $-$0&.03  \\
     6922 &   6.50 &   6.20 &   6.65 &   $-$7.48 &   3.00 &   2.50 & $-$0&.19  \\
     7990 &   6.50 &   6.35 &   6.84 &   $-$6.76 &   2.99 &   2.50 & $-$0&.31  \\
     8907 &   6.50 &   6.51 &   7.03 &   $-$6.66 &   3.00 &   2.50 & $-$0&.42  \\
     9412 &   6.50 &   6.54 &   6.95 &   $-$5.29 &   3.02 &   2.50 & $-$0&.49  \\
     9989 &   6.50 &   6.95 &   7.24 &   $-$6.75 &   3.02 &   2.50 & $-$0&.49  \\
    10500 &   6.50 &   6.99 &   7.24 &   $-$5.80 &   3.01 &   2.05 & $-$0&.32  \\
    11000 &   6.50 &   7.05 &   7.28 &   $-$5.77 &   3.01 &   2.05 & $-$0&.13  \\
    11499 &   6.50 &   7.09 &   7.33 &   $-$5.83 &   3.01 &   2.05 &   0&.12  \\
\noalign{\smallskip}
\hline
\noalign{\smallskip}
     6011$^{a}$
		      &   5.00 &   8.04 &   8.08 &   $-$7.93 &   8.38 &   5.00 &    1&.08  \\
\noalign{\smallskip}
\hline
\noalign{\smallskip}
\multicolumn{9}{@{}p{0.475\textwidth}@{}}{$^{a}$Stellar simulation with $N_{\rm He}/N_{\rm H}$ = 8.51\% and [M/H] = $-4.0$. \newline {\bf Notes.} All quantities were averaged over 12 snapshots. \Te\ is derived from
  the temporal and spatial average of the emergent flux, $x$ (same as $y$) and $z$
  correspond to the geometrical dimensions of the box, and $\tau_{\rm
  R}$ is the Rosseland optical depth averaged over constant
  geometrical depth. $t$ is the computation time in
  stellar time and $t_{\rm adv}$ is the advective or turnover timescale at the geometrical depth that
  corresponds to $\langle \tau_{\rm
  R} \rangle_{\rm x,y} = 1$.}
\end{tabular}
\label{tab:grid}
\end{center}
\end{table}

\begin{table}[!t]
\caption{Granulation Properties of ELM WDs}
\begin{center}
\setlength{\tabcolsep}{3.25pt}
\begin{tabular}{@{}rcccrrc@{}}
\hline
\hline
\noalign{\smallskip}
\multicolumn{1}{c}{\Te} & \logg         & $\log$ Char. Size & $\log H_{\rm p}$ & \multicolumn{1}{c}{$\log t_{\rm decay}$} & \multicolumn{1}{c}{$\delta I_{\rm rms}/\langle I \rangle$} & $\log$ Mach \\
\multicolumn{1}{c}{(K)} & (cm s$^{-2}$) & (cm)              & (cm)             & \multicolumn{1}{c}{(s)}                  & \multicolumn{1}{c}{(\%)}                                   &             \\
\noalign{\smallskip}
\hline
\noalign{\smallskip}
     5977 &   5.00 &   7.50 &   6.70 &   1.53 &  16.91 & $-$0.42\\
     6958 &   5.00 &   7.54 &   6.74 &   1.62 &  23.95 & $-$0.26\\
     7513 &   5.00 &   7.57 &   6.77 &   1.65 &  23.17 & $-$0.12\\
     8501 &   5.00 &   7.83 &   6.86 &   1.51 &  18.34 & $-$0.07\\
     9001 &   5.00 &   7.69 &   6.92 &   1.45 &  11.17 & $-$0.26\\
     9499 &   5.00 &   7.77 &   6.99 &   1.49 &   3.80 & $-$0.58\\
\noalign{\smallskip}
\hline
\noalign{\smallskip}
     6044 &   5.50 &   6.78 &   6.21 &   1.12 &  13.00 & $-$0.50\\
     6915 &   5.50 &   6.91 &   6.25 &   1.15 &  21.22 & $-$0.34\\
     7967 &   5.50 &   7.08 &   6.31 &   1.17 &  23.14 & $-$0.14\\
     8996 &   5.50 &   7.42 &   6.41 &   1.19 &  18.83 & $-$0.06\\
     9499 &   5.50 &   7.27 &   6.47 &   0.98 &  13.01 & $-$0.22\\
     9998 &   5.50 &   7.32 &   6.53 &   0.79 &   6.93 & $-$0.42\\
    10500 &   5.50 &   7.46 &   6.62 &   0.80 &   1.84 & $-$0.86\\
\noalign{\smallskip}
\hline
\noalign{\smallskip}
     5958 &   6.00 &   6.23 &   5.71 &   0.86 &   8.02 & $-$0.62\\
     6971 &   6.00 &   6.40 &   5.77 &   0.80 &  17.55 & $-$0.43\\
     7962 &   6.00 &   6.53 &   5.82 &   0.80 &  22.37 & $-$0.27\\
     8963 &   6.00 &   6.63 &   5.88 &   0.80 &  21.21 & $-$0.04\\
     9491 &   6.00 &   6.81 &   5.94 &   0.80 &  19.55 & $-$0.05\\
     9999 &   6.00 &   6.82 &   6.01 &   0.58 &  13.76 & $-$0.24\\
    10499 &   6.00 &   6.83 &   6.07 &   0.37 &   7.65 & $-$0.39\\
    11001 &   6.00 &   6.81 &   6.15 &   0.64 &   2.13 & $-$0.73\\
\noalign{\smallskip}
\hline
\noalign{\smallskip}
     5986 &   6.50 &   5.73 &   5.21 &   0.43 &   5.13 & $-$0.71\\
     6922 &   6.50 &   5.89 &   5.28 &   0.37 &  12.49 & $-$0.52\\
     7990 &   6.50 &   6.00 &   5.33 &   0.33 &  19.63 & $-$0.36\\
     8907 &   6.50 &   6.13 &   5.38 &   0.37 &  21.18 & $-$0.21\\
     9412 &   6.50 &   6.31 &   5.42 &   0.38 &  20.45 & $-$0.10\\
     9989 &   6.50 &   6.47 &   5.48 &   0.30 &  19.67 & $-$0.06\\
    10500 &   6.50 &   6.40 &   5.55 &   0.17 &  16.19 & $-$0.19\\
    11000 &   6.50 &   6.35 &   5.60 &   0.00 &   9.80 & $-$0.35\\
    11499 &   6.50 &   6.36 &   5.68 & $-$0.20 &   4.90 & $-$0.56\\
		\noalign{\smallskip}
\hline
\noalign{\smallskip}
     6011$^{a}$ 
		      &   5.00 &   7.34 &   6.60 &   1.76 &  17.79 & $-$0.38\\
\noalign{\smallskip}
\hline
\noalign{\smallskip}
\multicolumn{7}{@{}p{0.475\textwidth}@{}}{$^{a}$Stellar simulation with $N_{\rm He}/N_{\rm H}$ = 8.51\% and [M/H] = $-4.0$. \newline {\bf Notes.} All quantities were averaged over 250 snapshots except $T_{\rm eff}$ (see Table~1). The characteristic horizontal size of the granulation, 
	decay time, and relative intensity contrast ($\delta I_{\rm rms}/\langle I \rangle$)
  were computed from emergent intensity snapshots using the definitions given
  in \citet{tremblay13b}. Both the pressure scale height ($H_{\rm P}$) and the Mach number
  are computed at
  the geometrical depth that corresponds to $\langle \tau_{\rm
  R} \rangle_{\rm x,y} = 1$.}
\end{tabular}
\label{tab:gran}
\end{center}
\end{table}

The numerical setup for the ELM WD simulations is the same as
that described in \citet{tremblay13c} for a grid of
$150\times150\times150$ points. The properties of the boundary
conditions are reported in \citet[][see Sect.~3.2]{freytag12}. The
warmest simulations, represented by filled symbols in
Figure~\ref{fg:fHR}, have a convective zone that is thinner than the
vertical dimension of the atmosphere. In those cases, we use a bottom
layer that is closed to convective flows with imposed zero vertical
velocities. In cooler models, the convective flux is transported in
and out of the domain through an open boundary. In all cases the
lateral boundaries are periodic and the top boundary is open to
material flows and radiation. We use the same equation-of-state (EOS) 
and opacity tables as those described in \citet{tremblay13c}. 
In particular, we employ the \citet{hm88} EOS, the Stark
broadening profiles of \citet{TB09}, and the quasi-molecular line
opacity of \citet{allard04}. The wavelength-dependent opacities are
sorted in the 8 bin configuration discussed in \citet{tremblay13c}.


\subsection{Physical Properties}

\begin{figure}[t]
\centering
\includegraphics[scale=0.55,bb=60 380 473 700]{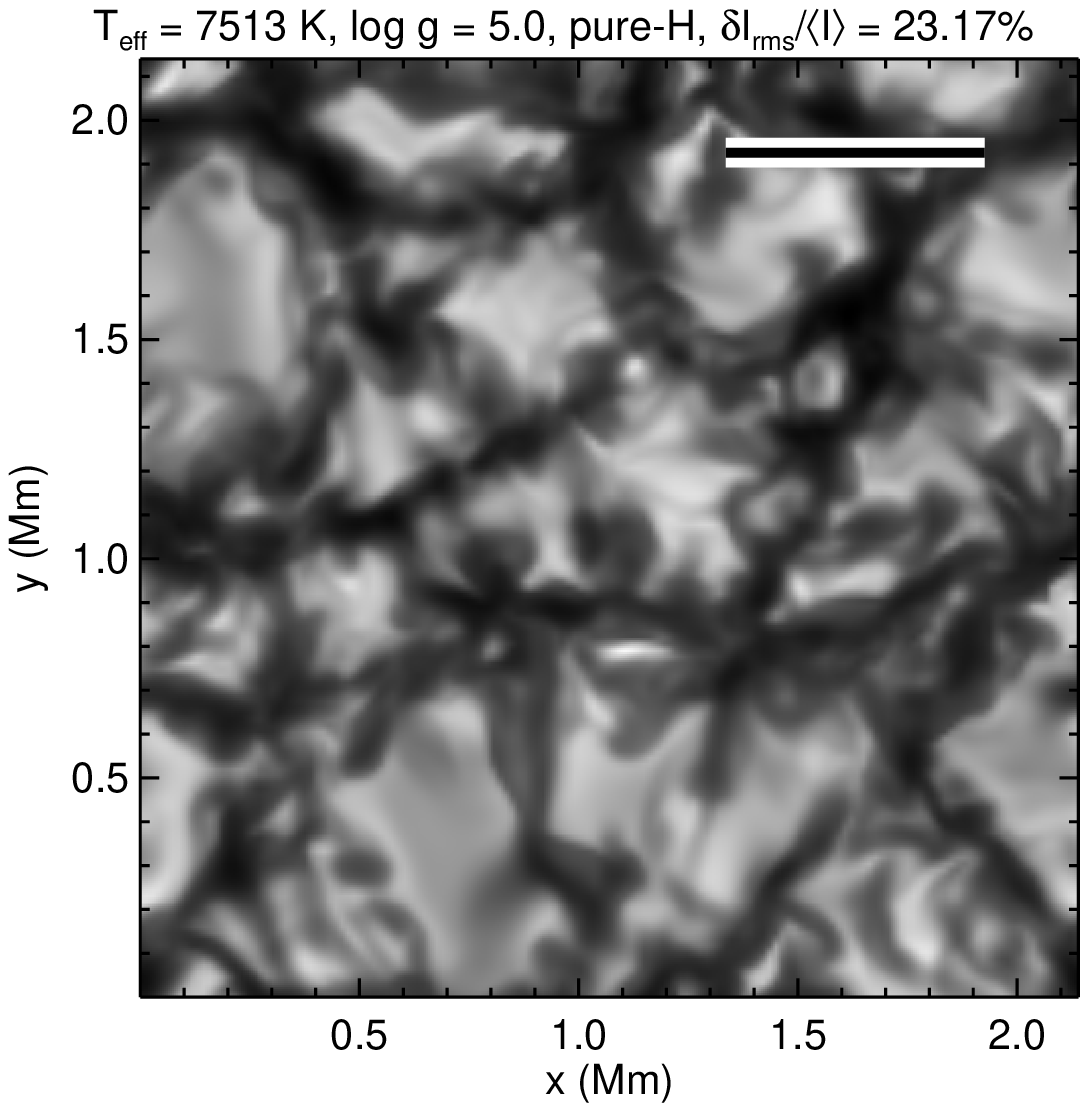}
\includegraphics[scale=0.55,bb=60 380 473 700]{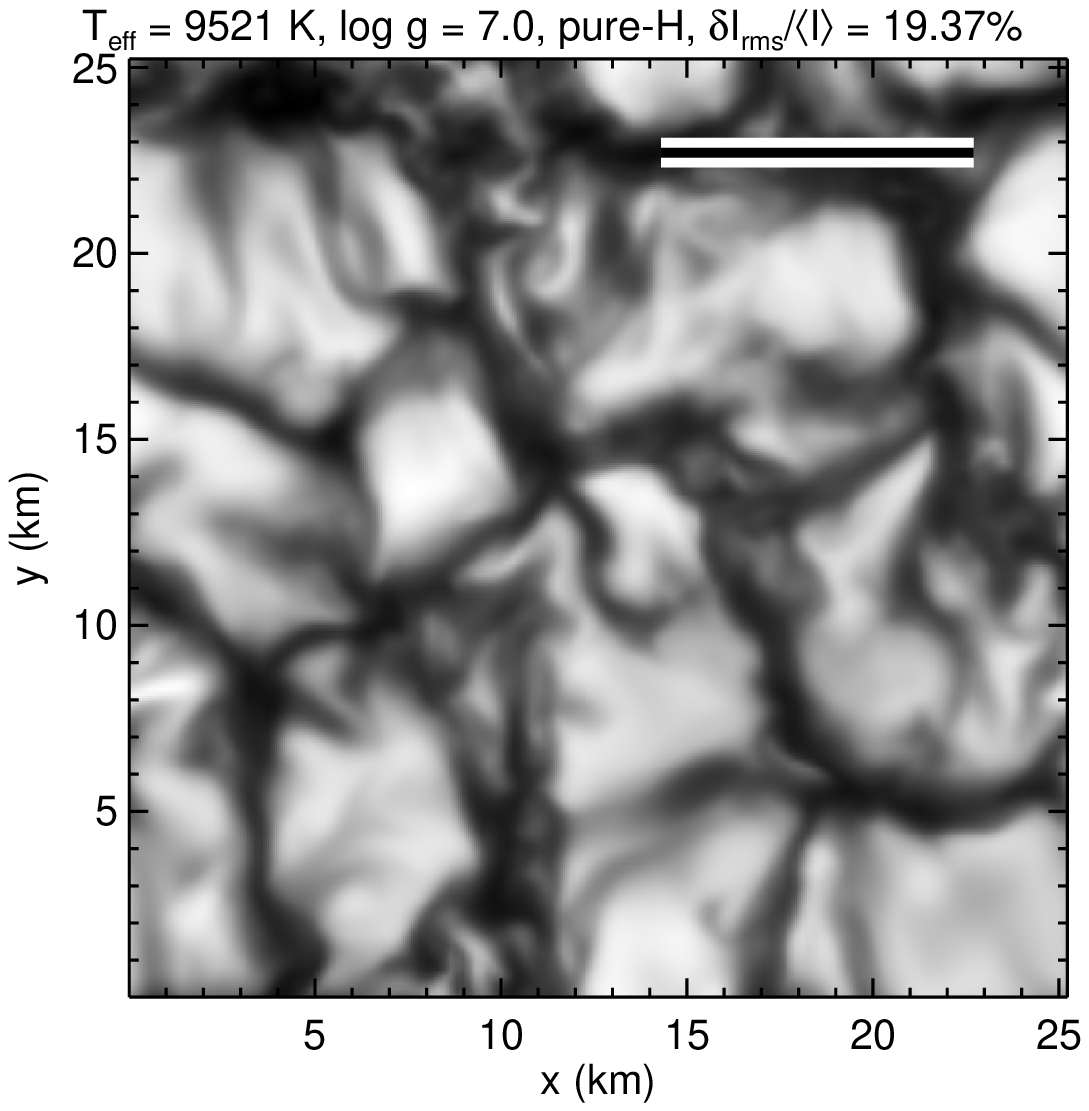}
\figcaption[d3t075g50wd08_xy.eps]{Emergent bolometric intensity at the
  top of the computational box for the pure-hydrogen WD
  simulations at \Te~=~7513~K, \logg~=~5.0 ({\it top}), and 9521~K,
  \logg~= 7.0 ({\it bottom}). The root-mean-square intensity contrast
  with respect to the mean intensity ($\delta I_{\rm rms}/\langle I
  \rangle$) is identified above the panels. The length of the bar in
  the top right is ten times the pressure scale height at $\langle
  \tau_{\rm R} \rangle_{\rm x,y} = 1$.  \label{fg:f3DA}}
\end{figure}

\begin{figure}[t]
\centering
\includegraphics[scale=0.55,bb=60 380 473 700]{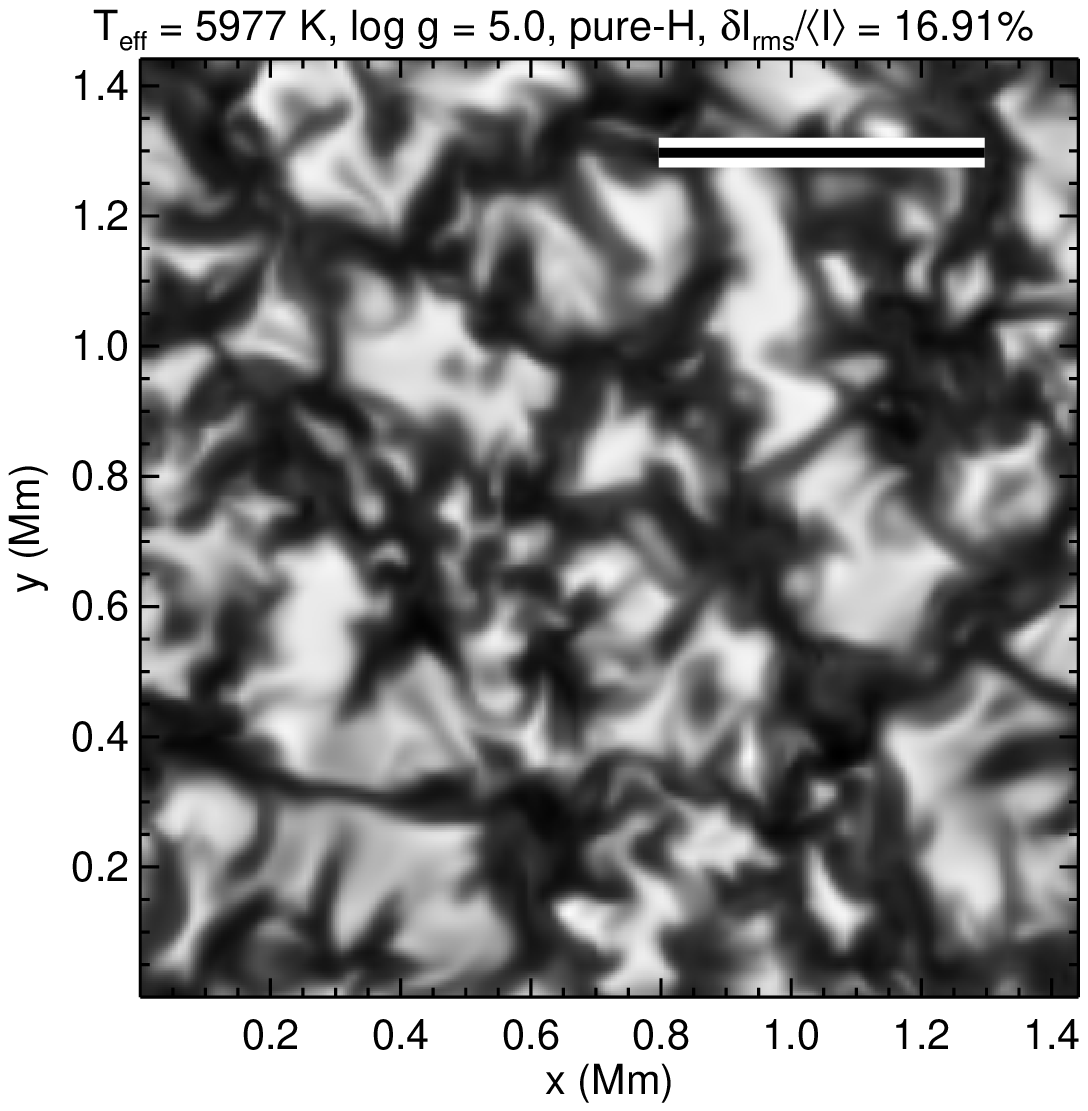}
\includegraphics[scale=0.55,bb=60 380 473 700]{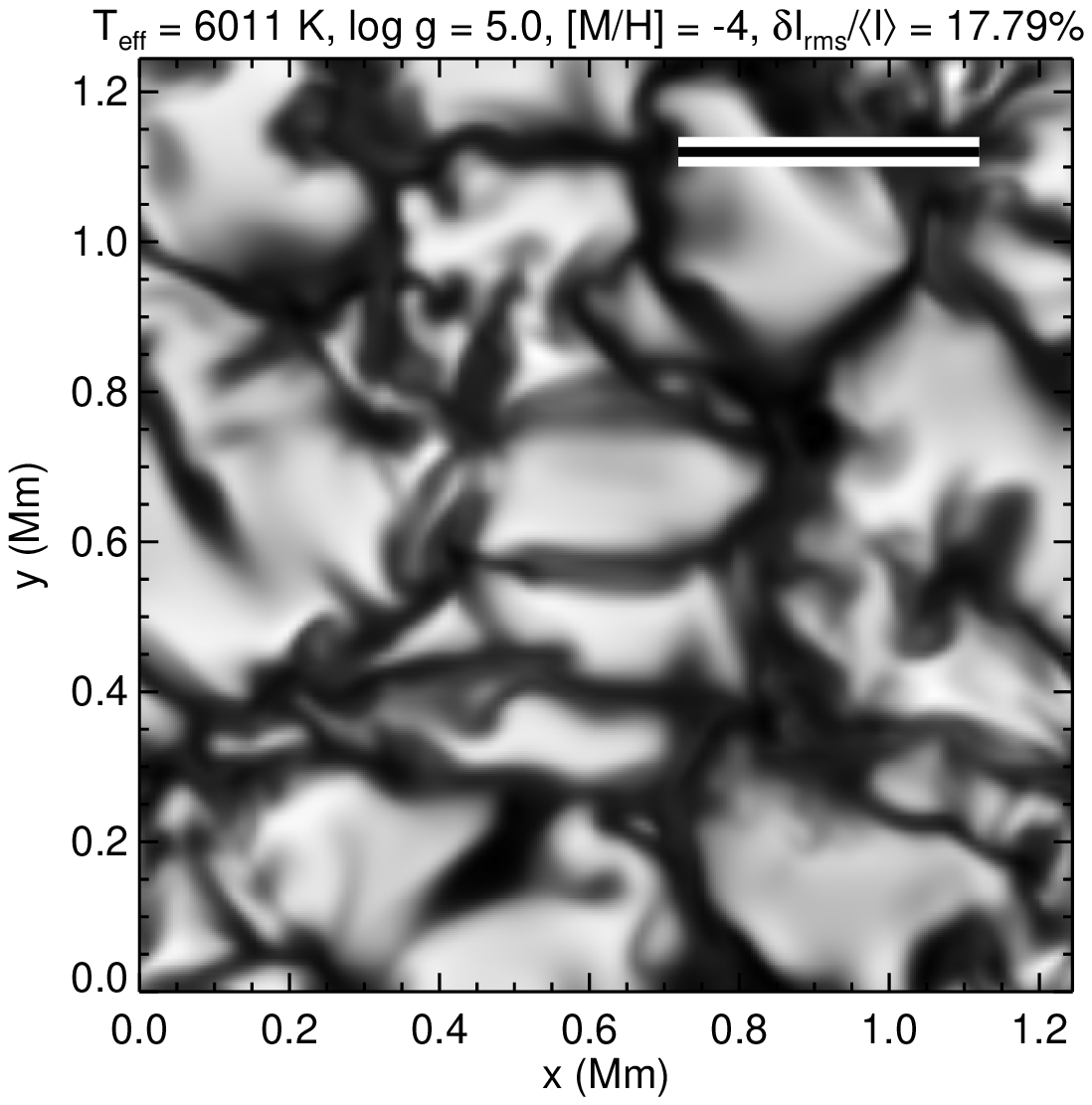}
\figcaption[d3t060g50wd08_xy.eps]{Similar to Figure~\ref{fg:f3DA} but
  for the emergent bolometric intensity of the pure-hydrogen WD 
	simulation at \Te~=~5977~K and \logg~=~5.0 ({\it top}), and
  the stellar simulation at 6011~K, \logg~=~5.0, and [M/H] = $-$4
  ({\it bottom}).\label{fg:f3DB}}
\end{figure}

\begin{figure}[t]
\centering
\includegraphics[scale=0.4]{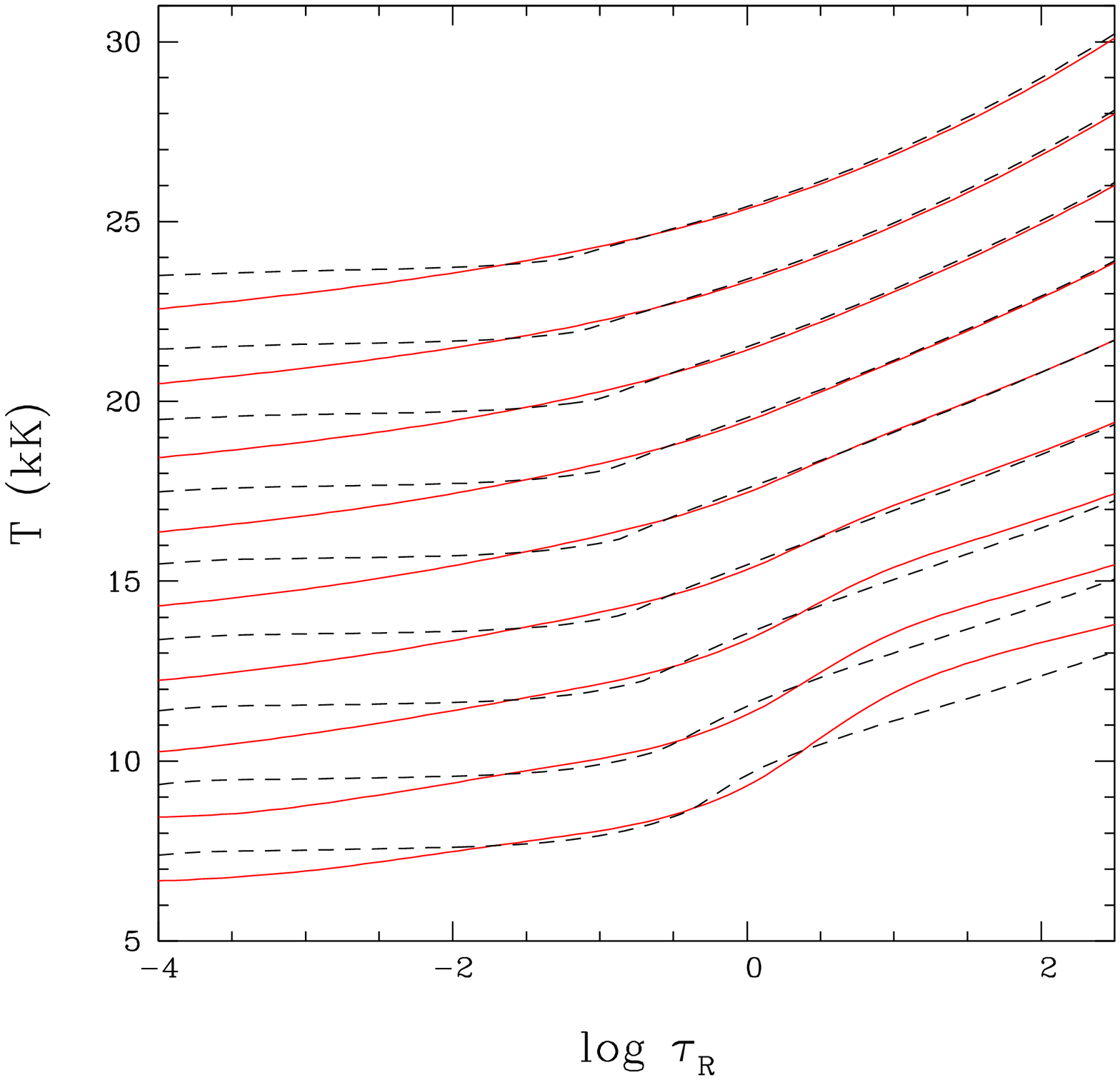}
\figcaption[f9.eps]{Temperature structure versus log $\tau_{\rm R}$
  for $\langle {\rm 3D}\rangle$ (red, solid) and 1D ML2/$\alpha$ = 0.8 (black, dashed)
  model atmospheres at $\sim$7000~K and gravities from \logg~=~5.0
  (bottom) to 9.0 (top) in steps of 0.5~dex. The temperature scale is
  correct for the \logg~=~5.0 model but other structures are shifted
  by 2~kK relative to each other for clarity. Properties of the 
	$\log g \geq 7.0$ models are given in \citet{tremblay13c}. 
\label{fg:ftstruct}}
\end{figure}

The ELM WD 3D simulations show physical properties that are
very similar to those of the 3D models of higher mass WDs presented in earlier
studies. \citet{tremblay13b} found that 3D effects on the thermal structure
are well correlated with the density at Rosseland optical depth $\tau_{\rm R} = 1$. Since the
atmospheric density can be kept roughly constant by decreasing both
\logg\ and \Te, the ELM simulations have properties similar to those
of C/O-core WDs but with a shift towards smaller \Te\ values.

Figure~\ref{fg:f3DA} compares our simulation at $\sim$7500~K and
\logg~=~5.0 with the simulation at $\sim$9500~K and \logg~=~7.0 from
\citet{tremblay13c}. Both simulations have a similar relative
intensity contrast and granulation is visually comparable. Table~\ref{tab:gran}
demonstrates that ELM WDs have a slightly larger maximum
relative intensity contrast compared to higher mass degenerates. ELM
WDs are very similar to main-sequence simulations in that
respect \citep[see Figure 10 of][]{tremblay13b}. Furthermore, Table~\ref{tab:gran} 
describes the characteristic size of granulation and decay time. These
quantities follow the same trends as those observed for higher mass WDs \citep{tremblay13b},
in particular the relation between the characteristic size of granulation to pressure
scale height ratio and photospheric Mach number.

Our low gravity WD simulations reach photospheric
plasma conditions and characteristic geometrical scales that
are similar to main-sequence stars, although the radii are quite different. In order to better understand
this link, we have computed a stellar model
at $T_{\rm eff} \sim$ 6000~K, $\log g = 5.0$, $N_{\rm He}/N_{\rm H}$ = 8.51\%, and a metallicity of [M/H] = $-$4.0
based on solar elemental abundances from \citet{grevesse98} with updated CNO values \citep{asplund05}.
The model is an extension of the CIFIST grid for 
main-sequence stars where $\log g \leq 4.5$. 
The simulation is representative of the extremely metal-poor F-dwarfs with 
the currently highest derived surface gravities \citep{spite12}. The properties
of the simulation are described in Tables~1 and 2, while Figure~\ref{fg:f3DB} 
shows the comparison with the pure-hydrogen ELM WD at 
\Te~$\sim$~6000~K and \logg~=~5.0. The relative intensity contrast
and size of the granules are similar, although the
stellar atmosphere is denser by 58\% at $\tau_{\rm R} = 1$ 
compared to the more opaque pure-hydrogen WD atmosphere. It implies
that their properties are only expected to be qualitatively
similar. Nevertheless, it demonstrates that ELM WDs
will provide an important connection between WDs and main-sequence stars
in future studies of the 3D effects on the determination of the atmospheric
parameters from Balmer line profiles.

We have computed mean 3D structures, hereafter $\langle {\rm
  3D}\rangle$ structures, which are spatial and temporal averages
of $T^4$ and $P$ performed over surfaces of constant $\tau_{\rm R}$ and for 12
random snapshots. We have employed those snapshots to define the $T_{\rm eff}$ values
of our simulations (Table~1). In Figure~\ref{fg:ftstruct}, we compare the
temperature structures of $\langle {\rm 3D}\rangle$ simulations and 1D
models at \Te~$\sim$~7000~K in the range $5.0 \leq \log g \leq
9.0$. Convection is still
fairly adiabatic in the photosphere for $\log g \geq 7.0$ in this \Te\ range, resulting in small differences
between the $\langle {\rm 3D}\rangle$ and 1D cases except for the
optically thin layers where 3D overshoot causes a significant dynamic
cooling. Density decreases for lower surface gravities and convection
becomes less efficient in the photosphere. Figure~\ref{fg:ftstruct}
demonstrates that differences between $\langle {\rm 3D}\rangle$ and 1D
structures become larger for ELM WDs at $T_{\rm eff} \sim$
7000~K. The observed pattern is very similar to the one at constant
$\log g$ and variable $T_{\rm eff}$ in Figure~7 of
\citet{tremblay13a}.

\section{3D ATMOSPHERIC PARAMETERS}

\begin{figure*}[t]
\centering
\includegraphics[angle=270,scale=0.55,bb=97 16 521 784]{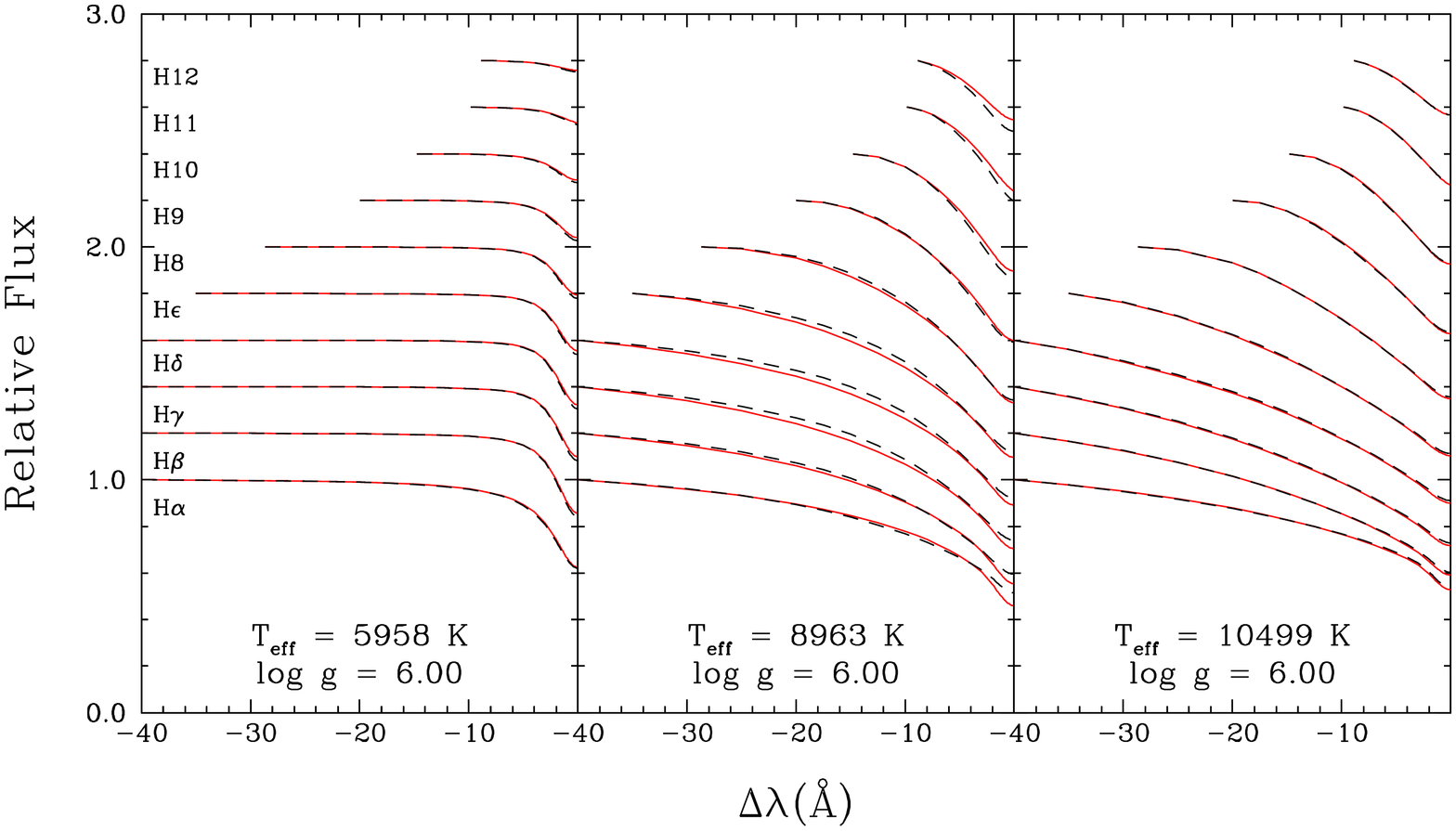}
\figcaption[f20b.eps]{Comparison of the blue wing of ten Balmer line
  profiles (H$\alpha$ to H12) calculated from $\langle {\rm
    3D}\rangle$ structures (red, solid) and standard 1D ML2/$\alpha$ = 0.8 structures
  (black, dashed) for three models at \logg~=~6.0 from our sequence
  identified in Table~\ref{tab:grid}. \Te\ values are shown on the different
  panels. All line profiles were normalized to a unity continuum at a
  fixed distance from the line center. The spectra were convolved with
  a Gaussian profile of 3~\AA\ resolution (FWHM) to represent typical
  observations.
\label{fg:fprofiles}}
\end{figure*}

\begin{figure}[h]
\centering
\includegraphics[scale=0.425,bb=20 167 592 654]{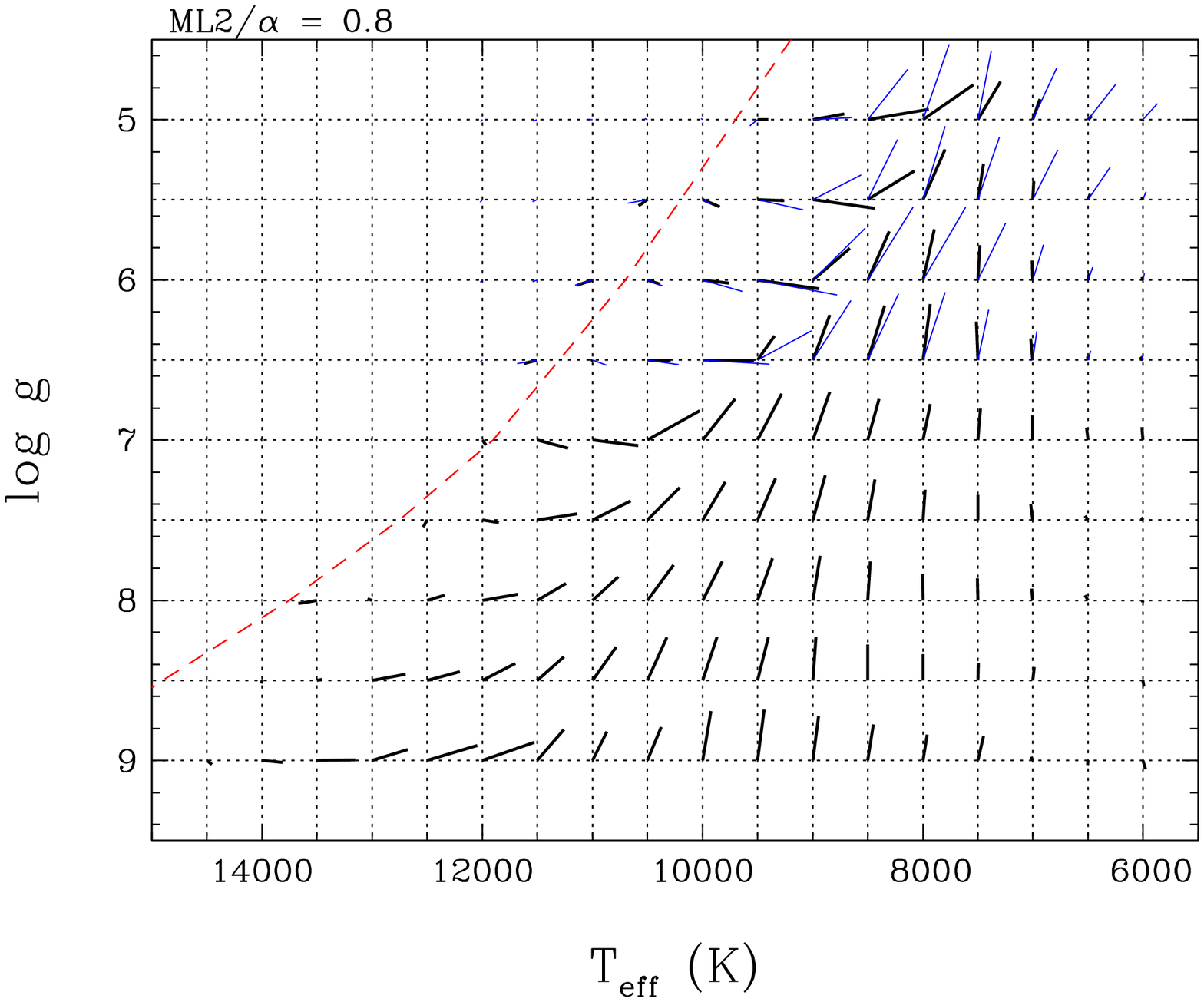}
\figcaption[f_shifts_all.eps]{$\langle$3D$\rangle$ atmospheric
  parameter corrections (solid black) found by fitting our grid of
  $\langle$3D$\rangle$ spectra with the reference grid of 1D spectra
  relying on the ML2/$\alpha$ = 0.8 parameterization of the MLT. The
  1D = 3D reference parameters are on the intersection of the dotted
  lines, and 3D corrections are read by following the solid lines from the intersection. We
  utilized a resolution of 3~\AA~and the cores of the deeper lines
  were removed from the fits \citep{tremblay13c}. The red dashed line
  represents the position of the maximum strength of H$\beta$ 
  for 1D models. Tabulated values are available in Table~\ref{tab:tcorr} 
  and \ref{tab:gcorr} for
  the range 5.0 $\leq \log g \leq$ 6.5 and in \citet{tremblay13c} for
  the range 7.0 $\leq \log g \leq$ 9.0. The thin solid blue lines
  represent 1.5D atmospheric parameter corrections (see Section 3.2).
\label{fg:fshifts}}
\end{figure}

\subsection{Mean 3D Spectra}

We have computed spectra from the 3D simulations in order to study
the 3D effects on the determination of atmospheric parameters. First of all, we
rely on mean spectra from $\langle {\rm 3D}\rangle$ structures,
hereafter $\langle {\rm 3D}\rangle$ spectra. For WD simulations with $\log g \geq 7.0$, 
the normalized Balmer line profiles computed from $\langle {\rm 3D}\rangle$
structures are nearly identical to the results of a full 3D spectral
synthesis \citep{tremblay11,tremblay13c}. However, we have found that the $\langle {\rm 3D}\rangle$ spectra
approximation becomes questionable at lower gravities, in line with
main-sequence simulations where 3D inhomogeneities have a significant
direct impact on predicted Balmer line profiles \citep{ludwig09}. We
review first the differences between $\langle {\rm 3D}\rangle$ and 1D spectra and
then consider the additional effects from 3D spectral synthesis in
Sect.~3.2.

Figure~\ref{fg:fprofiles} shows the comparison of predicted $\langle
{\rm 3D}\rangle$ and 1D Balmer line profiles for three simulations at $\log g
= 6.0$, which is an illustrative case for ELM WDs. The 1D
spectra adopt the ML2/$\alpha$ = 0.8 parameterization of the MLT
\citep{TB10} and the same microphysics as the 3D calculations.  The
ELM results can be compared to the case at $\log g = 8.0$ in Figure~16
of \citet{tremblay13a}. One major difference is that up to ten Balmer
lines are typically fitted simultaneously for ELM WDs, since
the non-ideal effects and the sensitivity to surface gravity are
shifted to higher Balmer lines \citep{gianninas14}. For the simulation at
$T_{\rm eff} \sim 9000$~K and $\log g = 6.0$ in the middle panel of
Figure~\ref{fg:fprofiles}, it is only for H10 and higher lines in the
series that $\langle {\rm 3D}\rangle$ profiles become shallower than
1D profiles due to differences in the predicted non-ideal effects.
The latter are very sensitive to the density stratification in the atmosphere
\citep{tremblay13c}. The shape of the lower Balmer lines is also
significantly different between $\langle {\rm 3D}\rangle$ and 1D
spectra at intermediate $T_{\rm eff}$, which is mostly responsible for
$T_{\rm eff}$ shifts.

The $\langle {\rm 3D}\rangle$ and 1D Balmer lines become 
identical at high $T_{\rm eff}$ values (see the right
panel of Figure~\ref{fg:fprofiles}). For ELM WDs, this transition
appears at relatively low temperatures, and 3D effects are largely
negligible on the thermal structures for $T_{\rm eff} > 10,000$~K.
Similarly to C/O-core WDs, convection becomes very inefficient 
in this regime and the thermal structure is essentially fixed by the radiation field, 
even though convective velocities are still large. For cool 
$T_{\rm eff}$ values, the left panel of Figure~\ref{fg:fprofiles} suggests
that 3D effects are small, but Section 3.2 demonstrates that
the $\langle {\rm 3D}\rangle$ approximation is not always good in this regime.

Figure~\ref{fg:fshifts} presents the $\langle {\rm 3D}\rangle$
atmospheric parameter corrections found by fitting the $\langle {\rm
  3D}\rangle$ spectra with our standard grid of 1D spectra. The
corrections were derived simultaneously for the nine Balmer lines from
H$\beta$ to H12 in the same way we fit observations. The $\langle {\rm
  3D}\rangle$ structures deviate significantly from their 1D
counterparts in the upper layers ($\tau_{\rm R} < 10^{-2}$) due to the
cooling effect of convective overshoot. This results in deep $\langle
{\rm 3D}\rangle$ cores for H$\alpha$ and H$\beta$, as observed in the
middle panel of Figure~\ref{fg:fprofiles}. While we have no suggestion
that the 3D simulations are inaccurate, there is no observational
evidence for an adiabatic surface cooling in WDs and it
appears premature to account for these effects. As a consequence, the
line cores were partially removed from the fits, as well as the entire
H$\alpha$ line, as discussed in Sect.~3.1 of \citet{tremblay13c}.

The $\langle {\rm 3D}\rangle$ atmospheric parameter corrections for
ELM WDs in Figure~\ref{fg:fshifts} follow the trend observed
for higher gravity objects, despite the fact that more lines are
included in the fitting procedure in the former case. The $\langle
{\rm 3D}\rangle$ $\log g$ corrections have a similar strength for ELM
and C/O-core WDs, although they are restricted to a narrower
range of $T_{\rm eff}$ at low surface gravity.

\subsection{3D Spectral Synthesis}

We have employed the Linfor3D three-dimensional spectral synthesis
code \citep{linfor} to compute 3D H$\beta$ profiles for ELM white
dwarfs. Unlike our previous experiments for $\log g \geq 7.0$, the 3D
spectra and their $\langle {\rm 3D}\rangle$ counterparts from properly
averaged $\langle {\rm 3D}\rangle$ structures differ by a few percent,
with the largest effects at lower gravities. It is currently out of
the scope of this work to proceed with the time-consuming computation
of a full grid of 3D synthetic spectra. However,
to further constrain the precision of the 3D corrections, we have
recomputed our grid of ELM spectra with the so-called 1.5D
approximation \citep{steffen95}. Under this formalism, we assume that
the physical conditions do not vary in the horizontal direction for
each point on the top of the computational box, i.e. each column is a 
plane-parallel atmosphere. The 1.5D spectrum is then the average
over all $150\times150$ columns, for which the emergent 1D spectra
can be easily calculated from our regular spectral synthesis code. The
full 3D spectral synthesis has the effect of coupling nearby grid
points, hence it is expected to lie somewhere between the extreme
cases of the 1.5D and $\langle {\rm 3D}\rangle$ approximations.

In Figure~\ref{fg:fshifts}, we compare the 1.5D (blue) and $\langle
{\rm 3D}\rangle$ (black) atmospheric parameter corrections. We notice
that while the $\langle {\rm 3D}\rangle$ approximation is relatively
good at $\log g = 6.5$, especially in terms of $\log g$ corrections, 
it becomes questionable at lower gravities. We remind the reader that the 1.5D approximation 
overestimates 3D effects, hence suggests an upper limit to the 3D corrections. Since both 1.5D and
$\langle {\rm 3D}\rangle$ corrections have the same signs, we suggest
that applying the $\langle {\rm 3D}\rangle$ corrections provides a
minimum 3D effect and a reasonable estimate of the atmospheric
parameters of ELM WDs.

The deficiency of the $\langle {\rm 3D}\rangle$ approximation for
ELM WDs is likely explained by multiple factors. The temperature
fluctuations are slightly larger at lower gravities although this
could not alone explain the variations. On the other hand, we 
find that the line to continuum emergent intensity ratio has a significantly larger
spatial variation for lower surface gravities. It implies
that fluctuations are less likely to cancel in the 3D spectral
synthesis. This could be related to the fact that broadening due to 
collisions with neutral atoms becomes increasingly important
at the low $T_{\rm eff}$ values where strong 3D effects are observed
in ELM WDs. Finally, we have verified that the
$\langle {\rm 3D}\rangle$ approximation is not the cause for the
slight residual bump in the 3D WD mass distribution at
$T_{\rm eff} \sim 12,000$~K \citep{tremblay13c,genest14} or the
presence of deep cores for the lower Balmer lines compared to 1D
spectra. These features are also present when using 1.5D spectra.


\subsection{3D Correction Functions}

\begin{table}[!t]
\small
\caption{1D ML2/$\alpha$ = 0.8 to 3D \Te\ Corrections}
\begin{center}
\begin{tabular}{@{\extracolsep{\fill}}rrrrr@{\extracolsep{\fill}}}
\hline
\hline
\noalign{\smallskip}
\multicolumn{1}{c}{\Te} & \logg~=~5.0    & \logg~=~5.5    & \logg~=~6.0   & \logg~=~6.5   \\
\multicolumn{1}{c}{(K)} & (cm s$^{-2}$) & (cm s$^{-2}$) & (cm s$^{-2}$) & (cm s$^{-2}$) \\
\noalign{\smallskip}
\hline
\noalign{\smallskip}
  6000 &      6 &     12 &     13 &     24 \\
  6500 &    $-$33 &    $-$19 &    $-$11 &      7 \\
  7000 &    $-$64 &     $-$6 &      5 &     21 \\
  7500 &   $-$203 &    $-$50 &    $-$19 &     16 \\
  8000 &   $-$460 &   $-$202 &   $-$105 &    $-$66 \\
  8500 &   $-$555 &   $-$425 &   $-$197 &   $-$152 \\
  9000 &   $-$288 &   $-$564 &   $-$338 &   $-$154 \\
  9500 &    $-$99 &   $-$243 &   $-$557 &   $-$151 \\
 10000 &      0 &   $-$155 &   $-$237 &   $-$466 \\
 10500 &      0 &     78 &   $-$116 &   $-$211 \\
 11000 &      0 &      0 &    136 &     13 \\
 11500 &      0 &      0 &      0 &    123 \\
\noalign{\smallskip}
\hline
\noalign{\smallskip}
\end{tabular}
\label{tab:tcorr}
\end{center}
\end{table}

We have derived functions to convert 1D atmospheric parameters of ELM WDs ($\log g
\leq$ 7.0) to
their 3D counterparts. We have ensured that the transition is smooth at $\log g
\sim$ 7.0 with the functions already presented in \citet{tremblay13c}, which are more
precise in the range 7.0 $\leq \log g \leq$ 9.0. Fortran 77 functions
are available in the online Appendix A. One can also rely on the data
from Tables~\ref{tab:tcorr} and \ref{tab:gcorr} to derive alternative functions or
request $\langle {\rm 3D}\rangle$ spectra to the authors. Our independent
variables are defined as

\begin{equation}
g_{\rm X} = \log g~{\rm [cm~s^{-2}]} - 8.0 ~,
\end{equation}

\begin{equation}
T_{\rm X} = \frac{T_{\rm eff}~{\rm [K]} - 10,000}{1000} ~,
\end{equation}

{\noindent}and the fitting functions for the ML2/$\alpha$ = 0.8
parameterization of the MLT are given below with the numerical
coefficients identified in Table~\ref{tab:coeff}.

\begin{equation}
\begin{split}
\frac{\Delta T_{\rm eff}}{1000} {\rm ~[ML2/}\alpha~{\rm = 0.8]} =~&a_{13}+\Big(a_{1}+a_{4}\exp\Big[(a_{5}+a_{7}\exp[a_{8}T_{\rm X} \\
                                                                  &+ a_{9}g_{\rm X}])T_{\rm X}+a_{6}g_{\rm X}\Big]\Big)\exp\Big((a_{2}+ \\
                                                                  &a_{10}\exp[a_{11}T_{\rm X}+a_{12}g_{\rm X}])T_{\rm X}+a_{3}g_{\rm X}\Big),
\end{split}
\end{equation}

\begin{equation}
\begin{split}
\Delta \log g {\rm ~[ML2/}\alpha~{\rm = 0.8]} =~&b_{13}+(b_{1}+b_{10}\exp[b_{11}T_{\rm X}+b_{12}g_{\rm X}]) \\
                                                &\exp\Big([b_{2}+b_{4}\exp(b_{5}T_{\rm X}+b_{6}g_{\rm X})]T_{\rm X} \\ 
                                                &+ [b_{3}+b_{7}\exp(b_{8}T_{\rm X}+b_{9}g_{\rm X})]g_{\rm X}\Big).
\end{split}
\end{equation}

\begin{table}[!t]
\small
\caption{1D ML2/$\alpha$ = 0.8 to 3D \logg\ Corrections}
\begin{center}
\begin{tabular}{@{\extracolsep{\fill}}rrrrr@{\extracolsep{\fill}}}
\hline
\hline
\noalign{\smallskip}
\multicolumn{1}{c}{\Te} & \logg~=~5.0    & \logg~=~5.5    & \logg~=~6.0   & \logg~=~6.5   \\
\multicolumn{1}{c}{(K)} & (cm s$^{-2}$) & (cm s$^{-2}$) & (cm s$^{-2}$) & (cm s$^{-2}$) \\
\noalign{\smallskip}
\hline
\noalign{\smallskip}
  6000 &    0.010 & $-$0.019 & $-$0.017 & $-$0.018 \\
  6500 & $-$0.023 & $-$0.037 & $-$0.040 & $-$0.024 \\
  7000 & $-$0.128 & $-$0.117 & $-$0.122 & $-$0.134 \\
  7500 & $-$0.239 & $-$0.225 & $-$0.215 & $-$0.239 \\
  8000 & $-$0.218 & $-$0.316 & $-$0.315 & $-$0.350 \\
  8500 & $-$0.064 & $-$0.181 & $-$0.302 & $-$0.340 \\
  9000 & $-$0.035 &    0.054 & $-$0.199 & $-$0.283 \\
  9500 &    0.001 &    0.006 &    0.056 & $-$0.150 \\
 10000 &    0.000 &    0.044 &    0.021 &    0.004 \\
 10500 &    0.000 &    0.037 &    0.027 &    0.002 \\
 11000 &    0.000 &    0.000 &    0.034 &    0.007 \\
 11500 &    0.000 &    0.000 &    0.000 &    0.024 \\
\noalign{\smallskip}
\hline
\noalign{\smallskip}
\end{tabular}
\label{tab:gcorr}
\end{center}
\end{table}

\begin{table}[!t]
\small
\caption{Coefficients for 3D Correction Functions}
\begin{center}
\begin{tabular}{@{\extracolsep{\fill}}crcr@{}}
\hline
\hline
\noalign{\smallskip}
Coefficient & \multicolumn{1}{c}{Value} & Coefficient & \multicolumn{1}{c}{Value} \\
(\Te)       &                           & (\logg)     &                           \\
\noalign{\smallskip}
\hline
\noalign{\smallskip}
a$_{1}$ & $-$7.3801851E+00        & b$_{1}$ & 5.1729169E$-$02     \\
a$_{2}$ & 6.0978389E$-$01         & b$_{2}$ & $-$2.8514123E+00    \\
a$_{3}$ & $-$8.5916775E$-$01      & b$_{3}$ & $-$2.8616686E+00    \\
a$_{4}$ & 7.2038240E+00           & b$_{4}$ & 3.2487645E+00       \\
a$_{5}$ & $-$1.1195542E$-$01      & b$_{5}$ & $-$7.3122419E$-$02  \\
a$_{6}$ & $-$8.2204863E$-$03      & b$_{6}$ & $-$3.5157643E$-$02  \\
a$_{7}$ & 1.0550418E$-$01         & b$_{7}$ & 2.0585494E+00       \\
a$_{8}$ & 2.2982103E$-$03         & b$_{8}$ & 2.9195288E$-$01     \\
a$_{9}$ & $-$2.3313912E$-$02      & b$_{9}$ & $-$9.2196599E$-$02  \\
a$_{10}$ & $-$1.5513621E$-$01     & b$_{10}$& $-$3.2042870E$-$01  \\
a$_{11}$ & 6.5429395E$-$01        & b$_{11}$& $-$6.3154984E$-$01  \\
a$_{12}$ & $-$1.1711317E+00       & b$_{12}$& 8.3885527E$-$01     \\
a$_{13}$ & 2.8245633E$-$03        & b$_{13}$& $-$1.0552187E$-$04  \\
\noalign{\smallskip}
\hline
\noalign{\smallskip}
\end{tabular}
\label{tab:coeff}
\end{center}
\end{table}

\section{ASTROPHYSICAL IMPLICATIONS}

\begin{table*}
\caption{Convective ELM WD Parameters after 3D Corrections}
\begin{center}
\begin{tabular*}{0.9\hsize}{@{\extracolsep{\fill}}lr@{ $\pm$ }@{\extracolsep{0pt}}lr@{ $\pm$ }@{\extracolsep{0pt}}lcr@{ $\pm$ }@{\extracolsep{0pt}}rr@{ $\pm$ }@{\extracolsep{0pt}}lcr@{ $\pm$ }@{\extracolsep{0pt}}lr@{ $\pm$ }@{\extracolsep{0pt}}lcc@{}}
\hline
\hline
\noalign{\smallskip}
Object & \multicolumn{2}{c}{\Te} & \multicolumn{2}{c}{\logg}         & $M$     & \multicolumn{2}{c}{$D$}  & \multicolumn{2}{c}{$R$}         & $\tau_{\rm cool}$ & \multicolumn{2}{c}{$M_{\rm 2, min}$} & \multicolumn{2}{c}{$M_{\rm 2, i=60^{\degree}}$} & $\tau^{\rm max}_{\rm merge}$ & $\log$~h \\
       & \multicolumn{2}{c}{(K)} & \multicolumn{2}{c}{(cm s$^{-2}$)} & (\msun) & \multicolumn{2}{c}{(pc)} & \multicolumn{2}{c}{{(0.01\rsun)}} & (Gyr)             & \multicolumn{2}{c}{(\msun)}          & \multicolumn{2}{c}{(\msun)}                     & (Gyr)                        &          \\
\noalign{\smallskip}
\hline	
\noalign{\smallskip}
J0112+1835   &  9740 & 140 & 5.77 & 0.05 & 0.160 &  662 &  98 & 8.63 & 1.07 & 1.8 & 0.62 & 0.03 & 0.85 & 0.04 & 2.68 & $-$22.40 \\
J0152+0749   & 10800 & 190 & 5.93 & 0.05 & 0.168 &  980 & 131 & 7.36 & 0.83 & 1.4 & 0.57 & 0.03 & 0.78 & 0.04 &\ldots& $-$22.80 \\
J0345+1748   &  8560 & 120 & 6.51 & 0.04 & 0.181 &  175 &  34 & 3.91 & 0.41 & 3.9 & 0.76 & 0.03 &\multicolumn{2}{c}{\ldots} & 7.18 & $-$22.16 \\
J0745+1949   &  8230 & 130 & 5.91 & 0.07 & 0.156 &  300 &  56 & 7.28 & 1.07 & 3.6 & 0.10 & 0.01 & 0.12 & 0.02 & 5.80 & $-$22.65 \\
J0818+3536   & 10190 & 190 & 5.87 & 0.09 & 0.164 & 3388 & 640 & 7.76 & 1.26 & 1.7 & 0.25 & 0.04 & 0.33 & 0.06 & 9.30 & $-$23.47 \\
J0822+2753   &  8800 & 130 & 6.36 & 0.05 & 0.175 &  496 &  72 & 4.57 & 0.51 & 4.2 & 0.76 & 0.08 & 1.05 & 0.11 & 8.18 & $-$22.31 \\
J0849+0445   & 10020 & 150 & 6.27 & 0.06 & 0.176 & 1101 & 158 & 5.08 & 0.61 & 1.9 & 0.65 & 0.04 & 0.89 & 0.05 & 0.45 & $-$22.38 \\
J0900+0234   &  8330 & 130 & 5.96 & 0.07 & 0.155 &  618 & 113 & 6.83 & 0.99 & 3.8 &\multicolumn{2}{c}{\ldots}&\multicolumn{2}{c}{\ldots}&\ldots&\ldots\\
J1005+3550   &  9760 & 140 & 6.00 & 0.05 & 0.166 & 1231 & 392 & 6.72 & 0.79 & 2.1 & 0.19 & 0.02 & 0.24 & 0.02 &10.63 & $-$23.12 \\
J1112+1117   &  9240 & 140 & 6.17 & 0.06 & 0.169 &  266 &  42 & 5.57 & 0.70 & 2.7 & 0.14 & 0.01 & 0.17 & 0.02 &12.73 & $-$22.56 \\
J1443+1509   &  8970 & 130 & 6.44 & 0.06 & 0.181 &  557 &  85 & 4.24 & 0.51 & 3.5 & 0.84 & 0.12 & 1.17 & 0.17 & 3.82 & $-$22.25 \\
J1512+2615   & 11250 & 180 & 6.93 & 0.06 & 0.250 &  766 &  96 & 2.85 & 0.31 & 0.4 & 0.31 & 0.04 & 0.39 & 0.05 &\ldots& $-$22.94 \\
J1518+0658   &  9650 & 140 & 6.68 & 0.05 & 0.197 &  309 &  41 & 3.38 & 0.35 & 0.9 & 0.58 & 0.03 & 0.78 & 0.04 &\ldots& $-$22.42 \\
J1538+0252   & 10030 & 150 & 5.98 & 0.06 & 0.167 & 1154 & 180 & 6.90 & 0.92 & 1.9 & 0.76 & 0.06 & 1.06 & 0.09 &\ldots& $-$22.86 \\
J1557+2823   & 12470 & 200 & 7.76 & 0.05 & 0.448 &  191 &  18 & 1.45 & 0.11 & 0.4 & 0.41 & 0.03 & 0.51 & 0.04 &\ldots& $-$21.92 \\
J1614+1912   &  8700 & 170 & 6.32 & 0.13 & 0.172 &  207 &  54 & 4.73 & 1.00 & 4.1 &\multicolumn{2}{c}{\ldots}&\multicolumn{2}{c}{\ldots}&\ldots&\ldots\\
J1630+2712   & 10200 & 170 & 6.03 & 0.08 & 0.169 & 2161 & 374 & 6.56 & 0.99 & 1.8 & 0.52 & 0.04 & 0.70 & 0.06 &\ldots& $-$23.13 \\
J1741+6526   & 10410 & 170 & 6.02 & 0.06 & 0.170 &  998 & 150 & 6.71 & 0.87 & 1.7 & 1.11 & 0.05 & 1.57 & 0.06 & 0.16 & $-$22.11 \\
J1840+6423   &  9120 & 140 & 6.34 & 0.05 & 0.177 &  740 & 109 & 4.70 & 0.54 & 3.3 & 0.65 & 0.03 & 0.88 & 0.04 & 4.78 & $-$22.47 \\
J2103$-$0027 &  9900 & 150 & 5.79 & 0.05 & 0.161 & 1253 & 177 & 8.47 & 1.01 & 1.8 & 0.71 & 0.04 & 0.98 & 0.05 & 5.73 & $-$22.72 \\
J2119$-$0018 &  9690 & 150 & 5.72 & 0.08 & 0.158 & 2835 & 507 & 9.09 & 1.38 & 1.7 & 0.74 & 0.04 & 1.03 & 0.05 & 0.58 & $-$22.82 \\
J2228+3623   &  7890 & 120 & 5.78 & 0.08 & 0.175 &  420 &  79 & 8.96 & 1.34 & 7.7 &\multicolumn{2}{c}{\ldots}&\multicolumn{2}{c}{\ldots}&\ldots&\ldots\\
J2236+2232   & 11310 & 170 & 6.54 & 0.04 & 0.182 &  355 &  43 & 3.81 & 0.40 & 1.1 & 0.39 & 0.03 & 0.51 & 0.04 &\ldots& $-$22.79 \\
\noalign{\smallskip}
\hline
\noalign{\smallskip}
\multicolumn{17}{@{}l}{{\bf Notes.} J0345+1748 is NLTT~11748 and J2236+2232 is LP~400-22.}
\end{tabular*}
\end{center}
\label{tab:elm}
\end{table*}

\subsection{The ELM WD Sample}

\citet{gianninas14} present a 1D spectroscopic analysis of 61 low-mass
WDs identified in the ELM Survey. All but five of these objects are in
confirmed binary systems with periods ranging from 12 min to about a
day.  The majority of these targets are warmer than 12,000 K, where
the 3D corrections are negligible. However, there are 23 convective
WDs that are in the right $T_{\rm eff}$ and $\log{g}$ range for 3D
corrections to be significant. Table~\ref{tab:elm} presents the updated
parameters for these 23 systems, including temperature, surface
gravity, mass, radius, cooling age, merger time, and their
gravitational wave strain. The corrections are largest at around 8000
K for low-mass WDs. Hence, the parameters for relatively cool objects
like J0345+1748 (NLTT 11748) have changed significantly. Since the 3D
corrections lower the estimated $\log{g}$, they also imply lower
masses and gravitational wave strains for these systems.

1D spectroscopic analysis of \citet{gianninas14} find $T_{\rm eff} =
8680 \pm 120$ K and $\log{g} = 6.83 \pm 0.04$ for NLTT
11748\footnote{The earlier analysis by \citet{kilic10b} found $\log{g}
  = 6.54 \pm 0.05$. However, this was based on older WD models without the improved Stark
  broadening profiles of \citet{TB09}.}.  The 1D to 3D corrections
from our models revise these to $T_{\rm eff} = 8560 \pm 120$ K and
$\log{g} = 6.51 \pm 0.04$, a change of 0.3 dex in surface gravity.
Using the \citet{althaus13} evolutionary models, the revised mass,
distance, and radius are $M=0.18 \pm 0.02 M_{\odot}$, $d = 175 \pm 34$
pc, and $R = 0.0391 \pm 0.0041 R_{\odot}$, respectively.  Our distance
estimate is in excellent agreement with the parallax measured at the
USNO, $\pi = 5.6 \pm 0.9$ mas. \citet{kaplan14} use high-quality
photometry of NLTT 11748 to constrain the same parameters from the
observed eclipses. Depending on the thickness of the surface hydrogen
layer of the secondary (more massive) WD, they derive $M_1=0.136-0.162
M_{\odot}$ and $R_1=0.0423-0.0433 R_{\odot}$.  Our mass and radius
measurements, including the 3D correction factors, are consistent with
their results within the errors.

\subsection{The Mass-Radius Relation}

\begin{figure}
\centering
\includegraphics[scale=0.425,bb=40 117 552 654]{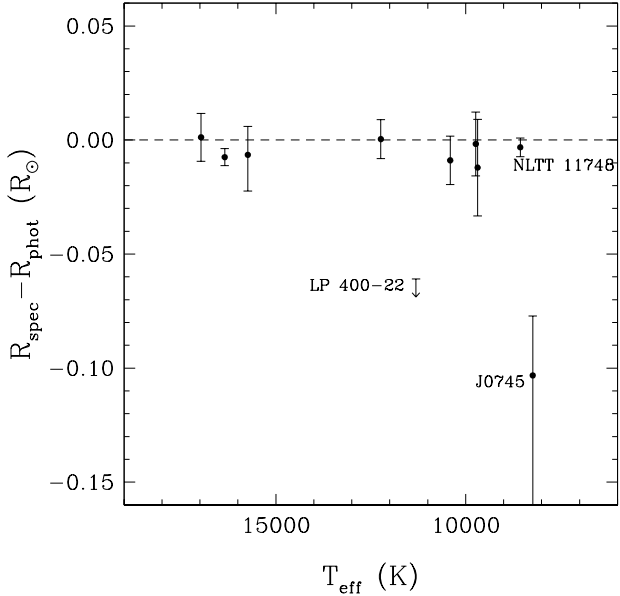}
\figcaption[plotdiff_R_3D.ps]{Differences in radii as measured from
  our 3D spectroscopic analysis and photometric analyses using
  eclipses or ellipsoidal variations. The error bars represent the
  errors of the two independent radius measurements added in
  quadrature. The errors for J0745$+$1949 reported in Figure 11 
	of \citet{gianninas14} are incorrect.
\label{fg:radius}}
\end{figure}

There are several tidally distorted and/or eclipsing ELM WD systems
that provide model-independent constraints on the radii. NLTT~11748 is
discussed above; it is an eclipsing system where the spectroscopically
inferred radius from the 1D models were significantly smaller than the
eclipse modelling suggested \citep{kaplan14,gianninas14}. Studying the
radius constraints on the ELM WD sample, \citet{gianninas14} find that
the agreement between the spectroscopically inferred radius and the
model-independent values is quite good for $T_{\rm eff} > 10,000$
K. However, there are four relatively cool objects, LP~400-22,
NLTT~11748, J0745, and J0751, where the radius is underestimated or the
mass/$\log{g}$ is overestimated.  Figure \ref{fg:radius} shows the comparison of the
updated radius measurements, including the 3D corrections, for the same
10 stars from \citet{gianninas14}. Table~\ref{tab:mr} provides the 
references for the different radius measurements. The agreement between the
spectroscopic and photometric radii is now excellent over
the temperature range 8,000--17,000~K. There are only two deviant
points, LP~400-22 and J0745, and we refer the reader to \citet{kilic13} 
and \citet{gianninas14b}, respectively, for further discussion of these objects. 
Ignoring LP~400-22 and J0745, the excellent agreement between the radius measurements from these two independent
methods suggests that the 3D model correction factors presented in
this paper have the right amplitude to account for the differences
seen in the earlier analysis relying on 1D models. Finally, we note that 
this independent verification of 3D results is unique to ELM WDs
and fairly different to the one presented in \citet{tremblay13c} 
for single C/O-core WDs with trigonometric parallax measurements.

\begin{table}[!t]
\caption{ELM WDs with Photometric Radius Measurements}
\begin{center}
\begin{tabular}{@{\extracolsep{\fill}}crrrc@{}}
\hline
\hline
\noalign{\smallskip}
Object & \multicolumn{1}{c}{\Te} & \multicolumn{1}{c}{$R_{\rm spec}$} & \multicolumn{1}{c}{$R_{\rm phot}$} & Ref. \\
       & \multicolumn{1}{c}{(K)} & \multicolumn{1}{c}{(0.01\rsun)}    & \multicolumn{1}{c}{(0.01\rsun)}    & \\
\noalign{\smallskip}
\hline
\noalign{\smallskip}
J0056$-$0611 & 12230 $\pm$ 180 & 5.65 $\pm$ 0.61 &   5.6  $\pm$ 0.6  & 1 \\
J0106$-$1000 & 16970 $\pm$ 260 & 6.42 $\pm$ 0.68 &   6.3  $\pm$ 0.8   & 1 \\
J0112$+$1835 & 9740  $\pm$ 140 & 8.64 $\pm$ 1.07 &   8.8  $\pm$ 0.9   & 1 \\
J0345$+$1748 & 8560  $\pm$ 120 & 3.91 $\pm$ 0.41 &   4.23 $\pm$ 0.10   & 2 \\
J0651$+$2844 & 16340 $\pm$ 260 & 3.25 $\pm$ 0.31 &   4.0  $\pm$ 0.2    & 1 \\
J0745$+$1949 & 8320  $\pm$ 130 & 7.28 $\pm$ 1.07 &  17.6$^{+9.0}_{-2.6}$   & 3 \\
J0751$-$0141 & 15740 $\pm$ 250 & 13.15 $\pm$ 1.04 &  13.8$^{+1.2}_{-0.7}$    & 1 \\
J1741$+$6526 & 10410 $\pm$ 170 & 6.71  $\pm$ 0.87 &   7.6  $\pm$ 0.6    & 1 \\
J2119$-$0018 & 9690  $\pm$ 150 & 9.09  $\pm$ 1.38 &  10.3  $\pm$ 1.6     & 1 \\
J2236$+$2232 & 11310 $\pm$ 170 & 3.81  $\pm$ 0.40 &   $>$9.9    & 4 \\
\noalign{\smallskip}
\hline
\noalign{\smallskip}
\multicolumn{5}{@{}p{0.475\textwidth}@{}}{{\bf References.} (1) \citet{hermes14}; (2) \citet{kaplan14}; (3) this work; (4) \citet{kilic13}. Spectroscopic parameters are taken from this work and \citet{gianninas14}.\newline
{\bf Notes.} See also \citet{gianninas14b} regarding the $R_{\rm phot}$ measurement uncertainties for J0745$+$1949.}
\end{tabular}
\label{tab:mr}
\end{center}
\end{table}

\subsection{ELM WDs with Milli-Second Pulsar Companions}

\begin{figure*}[!th]
\centering
\includegraphics[scale=0.41,bb=0 87 592 699]{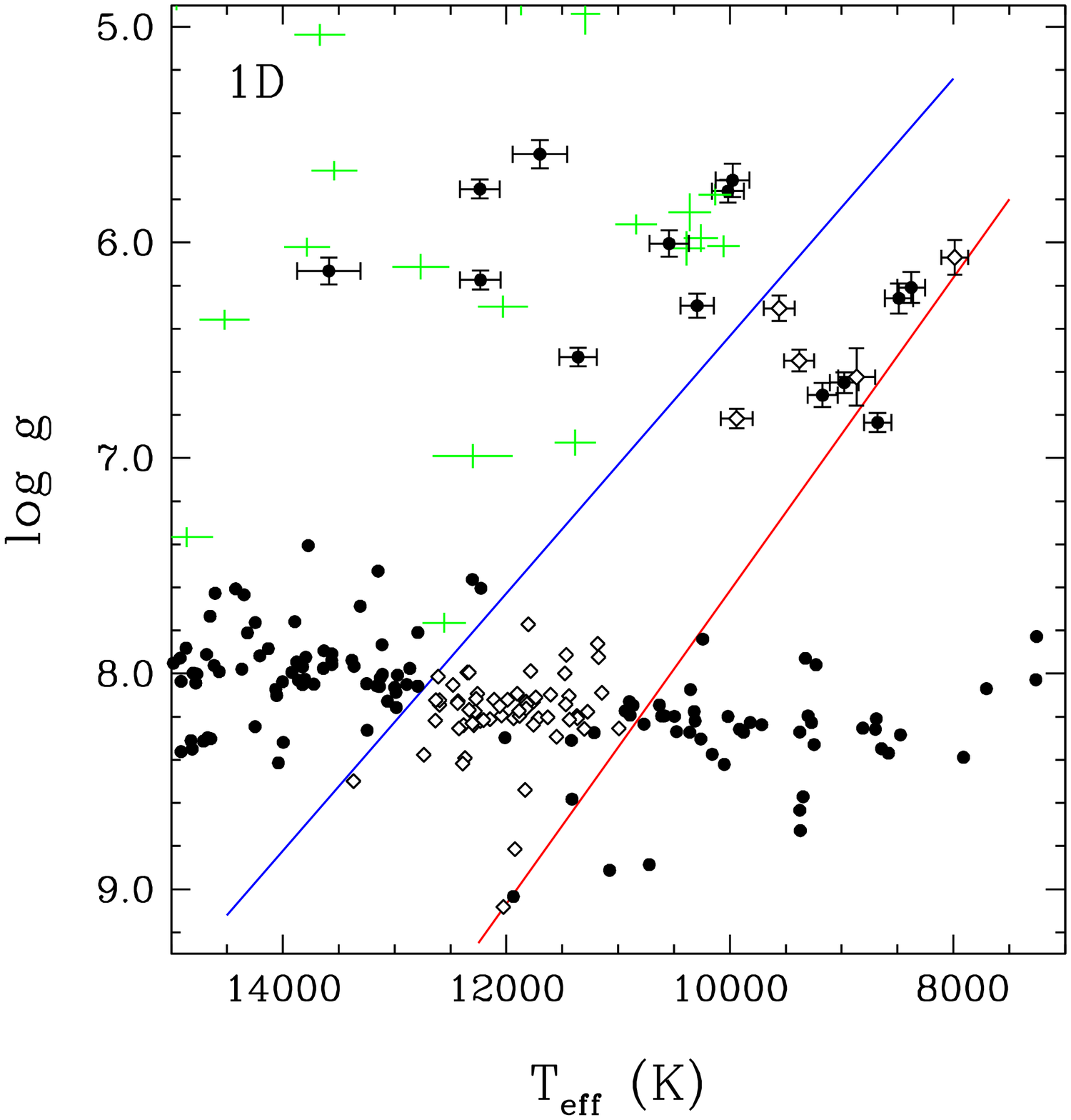}
\includegraphics[scale=0.41,bb=0 87 592 699]{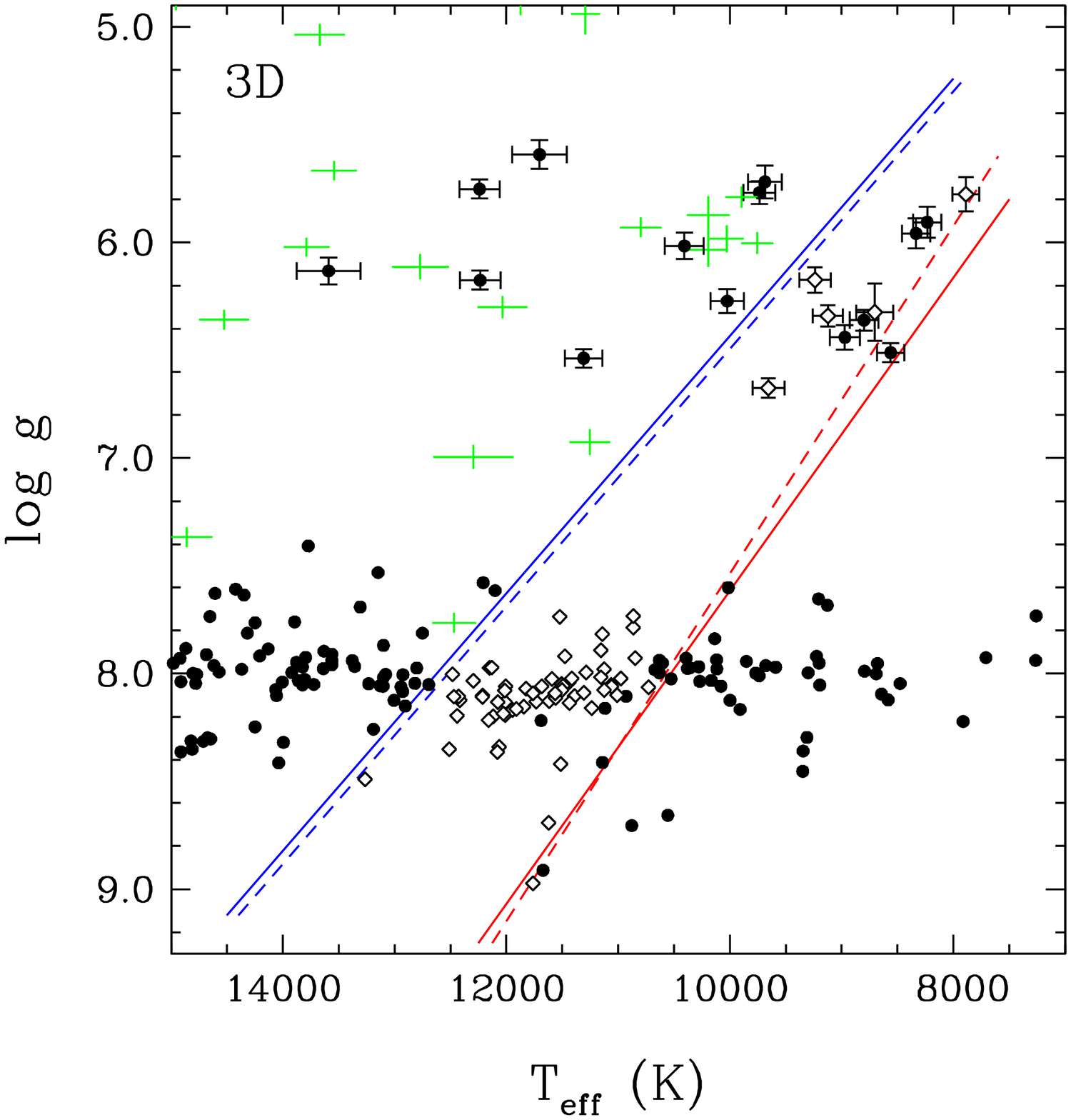}
\figcaption[ZZstrip_ELM_1D.ps]{The ZZ Ceti instability strip based on
  the 1D (left panel) and 3D (right panel) spectroscopic
  analysis. White diamonds represent the pulsators, whereas black dots
  show the photometrically constant WDs. The solid blue and red lines
  represent the empirical boundaries of the ZZ Ceti instability as
  determined by \citet{gianninas14}. The dashed lines denote tentative
  boundaries matching the location of both instability strips (for
  normal mass and ELM pulsators) based on the 3D corrected
  parameters. The green error bars denote the remaining ELM WDs that
  have not yet been investigated for photometric variability.
\label{fg:zz}}
\end{figure*}

There are several WD companions to milli-second pulsars with optical
spectroscopy available in the literature. These WDs provide an
independent method to estimate the mass of the neutron star through
constraints on the mass of the WD and the semi-amplitude of the radial
velocity variations. Hence, accurate mass (and $\log{g}$) measurements
for the ELM WDs in these systems are crucial.  Many of these WDs have
temperatures in the range 8000-10,000 K, where 3D effects are
significant.

PSR J0218+4232, J1012+5307, and J1909$-$3744 have companions with
$T_{\rm eff} \approx 8000-8500$ K and $\log{g}<7$, based on 1D
spectroscopic analyses
\citep{vankerkwijk96,callanan98,bassa03,jacoby05}.  The corrections in
$\log{g}$ are on the order of 0.3 dex at these temperatures. For
example, \citet{vankerkwijk96} derive $T_{\rm eff} = 8550 \pm 25$ K
and $\log{g} = 6.75 \pm 0.07$ for the ELM WD companion to PSR
J1012+5307. Taking the 3D corrections into account brings down these
values to $T_{\rm eff} = 8440$ K and $\log{g} = 6.43$. PSR
1911$-$5958A has a WD companion with $T_{\rm eff} = 10090 \pm 150$ K
and $\log{g} = 6.44 \pm 0.20$ \citep{bassa06}. The amplitude of the 3D
corrections is lower at these temperatures, bringing down the
temperature and surface gravity to 9780 K and $\log{g} = 6.38$,
respectively.

PSR J1738+0333 is an excellent system where the mass, radius, and the
surface gravity of the WD companion can be estimated from both optical
spectroscopy/photometry and radio observations. \citet{antoniadis12}
obtain $T_{\rm eff}= 9130 \pm 150$ K, $\log{g}= 6.55 \pm 0.10$,
$M=0.181^{+0.007}_{-0.005} M_{\odot}$, and $R=0.037^{+0.004}_{-0.003}
R_{\odot}$ from optical spectroscopy. Interestingly, they also derive
$\log{g}= 6.45 \pm 0.07$ (from the orbital period decay rate of the
binary, mass ratio, parallax, and photometry), $M=0.182 \pm 0.016
M_{\odot}$ (from the mass ratio and orbital decay rate of the binary),
and $R=0.042 \pm 0.004 R_{\odot}$ (from photometry). The latter
measurements are independent of the 1D spectroscopic analysis and
imply a lower surface gravity and a larger radius for this ELM
WD. These provide additional evidence for the ``high $\log{g}$
problem'' in 1D spectroscopic analysis of ELM WDs, though (admittedly)
the errors in surface gravity and radius are relatively large.  Based
on the 3D corrections factors that we estimate, we revise the
spectroscopic parameters of this companion to $T_{\rm eff} = 8910 \pm
150$ K and $\log{g} = 6.30 \pm 0.10$, which are consistent (within the
errors) with $\log{g}= 6.45 \pm 0.07$ estimated using the independent
method. We note that this WD companion also happens to be a pulsating WD
\citep{kilic15}.  

\subsection{The ZZ Ceti Instability Strip}

Figure \ref{fg:zz} shows the instability strip for the pulsating DAV WDs 
based on 1D and 3D spectroscopic analyses.  Here we only show the parameters for
the DAVs derived from a uniform analysis by
\citet{gianninas11,gianninas14}.  We also show the five ELM pulsators
currently known (excluding the companion to PSR J1738+0333 since the
1D model spectroscopic analysis of that star was not performed using
the same models), and a large number of photometrically constant stars
that help define the boundaries of the instability strip.  The solid
lines show the boundaries of the instability strip from Gianninas et
al. (2014).

There is a clear trend of increasing $\log{g}$ with decreasing
temperature in the 1D spectroscopic parameters of these stars. This
trend disappears in the 3D version of the plot, where the 3D
corrections lower the implied surface gravity. The blue edge of the
instability strip is well defined in the 1D analysis. The 3D
corrections move the stars systematically to slightly cooler
temperatures and lower surface gravities.  The dashed lines show the
revised boundaries based on the 3D corrected parameters. For the blue
edge, the boundary is simply shifted to the right by 100 K. The blue
edge of the instability strip is still well defined for both normal
mass and ELM WDs in the 3D analysis. For the red edge, as in the 1D
case, we use GD 518 and SDSS J222859.93+362359.6 as the two anchor
points at either extremity (the highest and lowest surface gravity) to
define the 3D red edge, which now has a higher slope. The equations 
of the revised blue and red edges are defined by Equations \ref{eq:blue} 
and \ref{eq:red}, respectively. Given the small number of ELM pulsators 
currently known, it is too early to definitively constrain the red edge 
of the instability strip for those stars. The new ZZ Ceti instability 
strip from the 3D analysis does not reveal any new ELM pulsator 
candidates in the published list of ELM WDs, but the revised boundaries 
should be used in the future to look for new pulsating DA WDs.

\begin{equation}
(\log g)_{\rm blue} = 5.96923 \times 10^{-4} (T_{\rm eff})_{\rm blue} + 0.52431
\label{eq:blue}
\end{equation}

\begin{equation}
(\log g)_{\rm red} = 8.06630 \times 10^{-4} (T_{\rm eff})_{\rm red} - 0.53039
\label{eq:red}
\end{equation}

\section{CONCLUSIONS}

We provide mean 3D model spectra for WDs with $5 \leq \log{g} < 7$,
extending the previously published 3D model grid to very low-mass
WDs. Our 3D models now cover the entire temperature and surface
gravity range for the observed population of convective DA WDs.  We list
1D to 3D correction factors, which can be as large as 0.35 dex in
surface gravity.  These corrections largely resolve the discrepancies seen in
the $\log{g}$ and radius estimates for relatively cool ELM WDs that are
observed in double WD or WD + milli-second pulsar systems. We also
update the boundaries of the ZZ Ceti instability strip for WDs,
including the ELM pulsators as well as the normal ($\log{g}\sim8$)
WDs.

\acknowledgements

Support for this work was provided by NASA through Hubble Fellowship
grant \#HF-51329.01 awarded by the Space Telescope Science Institute,
which is operated by the Association of Universities for Research in
Astronomy, Inc., for NASA, under contract NAS 5-26555.  MK and AG
gratefully acknowledge the support of the NSF and NASA under grants
AST-1312678 and NNX14AF65G.  AG acknowledges support provided by NASA 
through grant number HST-GO-13319.01 from the Space Telescope Science 
Institute, which is operated by AURA, Inc., under NASA contract 
NAS~5-26555. The computing for this project was performed at the OU 
Supercomputing Center for Education \& Research (OSCER) at the 
University of Oklahoma (OU). H.G.L. acknowledges financial support by 
the Sonderforschungsbereich SFB 881 ``The Milky Way System'' 
(subproject A4) of the German Research Foundation (DFG). J.J.H. acknowledges
funding from the European Research Council under the European Union's
Seventh Framework Programme (FP/2007-2013) / ERC Grant Agreement n.
320964 (WD-Tracer).

\clearpage
\include{tabelm}
\clearpage

\newpage

\appendix

\section{Correction functions for Fortran 77}

\lstset{
    breaklines     = true,
    numbers        = left,
    stepnumber     = 1,
}
\lstset{language=Fortran}

\subsection{1D ML2/$\alpha$ = 0.8 to 3D $T_{\rm eff}$ corrections}
\begin{lstlisting}
      function ML18_to_3D_dTeff_ELM(Teff,logg)
      real*8 ML18_to_3D_dTeff_ELM,a(13),Teff,logg
      real*8 TX,GX,Shift
c     ;
c     ; IN: Teff, effective temperature (K)
c     ; logg, surface gravity (g in cm/s2)
c     ;
c     ; OUT: Teff correction (K) 1D ML2/alpha = 0.8 to 3D
c     ; 
     
      A(1)= -7.3801851E+00
      A(2)=  6.0978389E-01
      A(3)= -8.5916775E-01
      A(4)=  7.2038240E+00
      A(5)= -1.1195542E-01
      A(6)= -8.2204863E-03
      A(7)=  1.0550418E-01
      A(8)=  2.2982103E-03
      A(9)= -2.3313912E-02
      A(10)=-1.5513621E-01
      A(11)= 6.5429395E-01
      A(12)=-1.1711317E+00
      A(13)= 2.8245633E-03

      TX=(Teff-10000.0)/1000.00
      GX=(logg-8.0)
      Shift=A(13)+(A(1)+A(4)*exp((A(5)+A(7)*exp(A(8)
     *     *TX+A(9)*GX))*TX+A(6)*GX))*exp((A(2)
     *      +A(10)*exp(A(11)*TX+A(12)*GX))*TX+A(3)*GX)

      ML18_to_3D_dTeff_ELM=Shift*1000.0
      return
      end 
\end{lstlisting}
~\\

\subsection{1D ML2/$\alpha$ = 0.8 to 3D $\log g$ corrections}

\begin{lstlisting}
      function ML18_to_3D_dlogg_ELM(Teff,logg)
      real*8 ML18_to_3D_dlogg_ELM,B(13),Teff,logg
      real*8 TX,GX,Shift
c     ;
c     ; IN: Teff, effective temperature (K)
c     ; logg, surface gravity (g in cm/s2)
c     ;
c     ; OUT: log g correction (g in cm/s2) 1D ML2/alpha = 0.8 to 3D
c     ; 

      B(1)=  5.1729169E-02
      B(2)= -2.8514123E+00
      B(3)= -2.8616686E+00
      B(4)=  3.2487645E+00
      B(5)= -7.3122419E-02
      B(6)= -3.5157643E-02
      B(7)=  2.0585494E+00
      B(8)=  2.9195288E-01
      B(9)= -9.2196599E-02
      B(10)=-3.2042870E-01
      B(11)=-6.3154984E-01
      B(12)= 8.3885527E-01
      B(13)=-1.0552187E-04
      
      TX=(Teff-10000.0)/1000.0
      GX=(logg-8.0)
      Shift=B(13)+(B(1)+B(10)*exp(B(11)*TX+B(12)*GX))
     *      *exp((B(2)+B(4)*exp(B(5)*TX+B(6)*GX))
     *      *TX+(B(3)+B(7)*exp(B(8)*TX+B(9)*GX))*GX)

      ML18_to_3D_dlogg_ELM=Shift
      return
      end
\end{lstlisting}
~\\

\end{document}